\documentclass[prb,twocolumn,showpacs,preprintnumbers,amsmath,amssymb]
{revtex4-1}
\usepackage{graphicx}
\usepackage{dcolumn}
\usepackage{bm}
\usepackage{mathrsfs}
\usepackage{subfigure}
\usepackage{natbib}
\usepackage{color}

\begin{document}

\title{Superconducting transition temperature of Pb nanofilms: Impact of the
thickness-dependent oscillations of the phonon mediated electron-electron
coupling} 


\author{Yajiang Chen}
\affiliation{Departement Fysica, Universiteit Antwerpen,
Groenenborgerlaan 171, B-2020 Antwerpen, Belgium}
\author{A. A. Shanenko}
\affiliation{Departement Fysica, Universiteit Antwerpen,
Groenenborgerlaan 171, B-2020 Antwerpen, Belgium}
\author{F. M. Peeters}
\email{francois.peeters@ua.ac.be}
\affiliation{Departement Fysica, Universiteit Antwerpen,
Groenenborgerlaan 171, B-2020 Antwerpen, Belgium}
\email{francois.peeters@ua.ac.be}
\date{\today}

\begin{abstract}
To date, several experimental groups reported measurements of the thickness
dependence of $T_c$ of atomically uniform single-crystalline ${\rm Pb}$
nanofilms. The reported amplitude of the $T_c$-oscillations varies significantly
from one experiment to another. Here we propose that the reason for this
unresolved issue is an interplay of the quantum-size variations in the
single-electron density of states with thickness-dependent oscillations in the
phonon mediated electron-electron coupling. Such oscillations in the coupling
depend on the substrate material, the quality of the interface, the protection
cover and other details of the fabrication process, changing from one experiment
to another. This explains why the available data do not exhibit one-voice
consistency about the amplitude of the $T_c$-oscillations. Our analyses are
based on a numerical solution of the Bogoliubov-de Gennes equations for a
superconducting slab.
\end{abstract}

\maketitle

\section{Introduction}
\label{introduction}
Quantum-size effects on the excitation
gap~\cite{PhysRevLett.10.332,PhysRevB.75.014519} and the transition
temperature~\cite{nature2_173, PhysRevLett.96.027005,Science316.1594} in
superconducting ultrathin films have been a subject of study since Blatt and
Thompson in 1963 predicted that the superconducting gap in ultrathin films
oscillates as a function of the thickness.~\cite{PhysRevLett.10.332} Such
oscillations are caused by the quantization of the electron motion in the
direction normal to the film. In particular, the formation of discrete
electron levels, i.e., the quantum well states
(QWS),~\cite{science12_1709,science324_1314} splits the three-dimensional (3D)
conduction band into a series of two-dimensional (2D) subbands. The bottom of
each subband is at the energy position of the corresponding QWS and so it is
shifted down (up) when increasing (decreasing) the nanofilm thickness. When one
of the QWS approaches the Fermi level, the density of the single-electron states
at the Fermi level $N(0)$ increases, resulting in enhanced superconductivity.

Recently, superconducting ${\rm Pb}$ nanofilms have been intensively
investigated experimentally where electronic effects are much stronger than
effects due to strain. The critical temperature of atomically flat ${\rm Pb}$
nanofilms with thickness down to a few monolayers on a ${\rm Si}(111)$ substrate
have been measured under different conditions.~\cite{science306_1915,
nature2_173, PhysRevLett.96.027005, Science316.1594,PhysRevLett.102.207002,
science324_1314, natphy6_104} It was found that superconductivity is not
destroyed by fluctuations even when the thickness of the nanofilm is only a
single monolayer,~\cite{natphy6_104} and the dependence of the critical
temperature on thickness shows oscillations with a period close to two
monolayers (ML),~\cite{science306_1915,PhysRevLett.96.027005} i.e., the even-odd
staggering in the superconducting properties.

In addition to $N(0)$, the phonon-mediated coupling $g$ between electrons also
makes an important contribution to superconductivity. Large
efforts~\cite{PhysRevLett.88.256802, PhysRevB.66.233408, PhysRevLett.95.096802,
PhysRevB.81.214519} have been made to understand how the formation of QWS can
affect the electron mass-enhancement factor $\lambda \sim gN(0)$~[only in the
weak-coupling limit $\lambda=gN(0)$]. Experimentally, from the information of
the temperature dependence of the line widths of QWS in ${\rm Ag}$ films
deposited on top of ${\rm Fe}(111)$ whisker, it was found that $\lambda$
exhibits an overall decrease with increasing film thickness so that
$\lambda$ approaches its bulk value as $\approx 1/N$, where $N$ is the number of
monolayers.~\cite{PhysRevLett.88.256802} The experimental study of $\lambda$ in
${\rm Pb}(111)$ nanofilms on silicon showed that the mass-enhancement factor
increases with increasing $N$. In both cases $\lambda$ exhibits pronounced
thickness-dependent oscillations. As it is mentioned above, a well-known reason
for these oscillations is the variation of $N(0)$ with the nanofilm thickness.
However, this is not the only important reason. There is an additional (less
known) effect coming from thickness-dependent oscillations of the coupling $g$.
The interplay of the oscillations of $g$ and $N(0)$ is not well understood yet
and often overlooked. In our previous paper~\cite{PhysRevB.75.014519} we
initiated a study of such an interplay based on intuitive arguments and used an
ansatz for the thickness dependent $g$. In the present paper we revisit the
problem and develop a simple model to justify this ansatz and to analyze all
the currently available experimental data of $T_c$ versus film thickness.
Our investigation is based on a numerical solution of the Bogoliubov-de Gennes
(BdG) equations. However, instead of solving the full BdG
formalism,~\cite{PhysRevB.75.014519} we make use of the Anderson recipe for a
very accurate approximate semi-analytical solution to the BdG equations:
corrections to the Anderson approximation are less than a few percent for
nanofilms. This will significantly simplify our numerical analysis.

Our paper is organized as follows. In Sec.~\ref{formalism} we outline the BdG
equations for a superconducting slab, including the basic moments concerning the
Anderson approximation. In Sec.~\ref{result} we discuss our numerical results on
the critical temperature in ultrathin superconducting nanofilms and highlight
important aspects due to the interplay of the thickness-dependent oscillations
of $N(0)$ and $g$. We demonstrate that this interplay is responsible for
significant variations in the amplitude of the quantum-size oscillations of
$T_c$ from one experiment to another. Our conclusions are given in
Sec.~\ref{conclusions}.

\section{Formalism}
\label{formalism}
\subsection{Bogolibov-de Gennes equations for a superconducting slab}
\label{formA}

The translational invariance is broken in the presence of quantum confinement so
that the superconducting order parameter $\Delta$ becomes position dependent in
nanoscale superconductors, i.e., $\Delta=\Delta({\bf r})$. In this case the
profile of $\Delta({\bf r})$ can be found from the BdG equations (see the
textbook~\cite{degen}) which are written as (for zero magnetic field)
\begin{eqnarray}\label{bdg}
\begin{bmatrix}  \hat{H}_{e} & \Delta({\bf r}) \\ \Delta({\bf r}) &- \hat{H}_{e}
 \end{bmatrix}\begin{bmatrix}u_{i{\bf k}}({\bf r}) \\ v_{i{\bf k}}({\bf
r})\end{bmatrix}=E_{ik}\begin{bmatrix}u_{i{\bf k}}({\bf r})\\ v_{i{\bf k}}({\bf
r})\end{bmatrix},
\end{eqnarray}
where $\hat{H}_{e}=-\frac{\hbar^2}{2m_{e}}\nabla^2-\mu$ is the single-electron
Hamiltonian (for zero magnetic field) measured from the chemical potential
$\mu$; $i$ is the quantum number associated with QWS for electron motion in
the $z$ direction perpendicular to the film, and ${\bf k}=(k_{x},k_{y})$
is the wavevector for quasi-free motion of electrons parallel to the film
~($k=|{\bf k}|$); $u_{i{\bf k}}({\bf r})$ and $v_{i{\bf k}}({\bf r})$ are the
particle- and hole-like wave functions; $E_{ik}$ stands for the quasi-particle
energy; and $\Delta({\bf r})$ is the superconducting order parameter (chosen as
real). The BdG equations are solved in a self-consistent manner with the
self-consistency condition given by~\cite{degen}
\begin{equation}\label{op}
\Delta({\bf r}) = g\sum_{i{\bf k}}u_{i{\bf k}}({\bf r})v^{*}_{i{\bf k}}({\bf
r})\tanh(\beta E_{ik}/2),
\end{equation}
where $g > 0$ is the effective coupling of electrons and $\beta=1/(k_BT)$. The
summation in Eq.~(\ref{op}) runs over states with the single-electron energy
$\xi_{ik}$ inside the Debye window, i.e., $|\xi_{ik}| < \hbar\omega_D$, with
$\hbar\omega_D=8.27\,{\rm meV}$ for ${\rm Pb}$~(see the
textbooks.~\cite{degen,fett})

As known, the BdG equations can, in principle, be solved without additional
approximations. However, even for ultrathin nanofilms this can be a
time-consuming task, especially in the vicinity of $T_c$, where a large number
of iterations is usually needed for a proper convergence. This is why it is
useful to invoke Anderson's recipe for a semi-analytical approximate solution.
The use of Anderson's recipe is well justified because (i) corrections were
found to be less than a few percent for nanofilms (see, e.g.,
Ref.~\onlinecite{PhysRevB.81.134523}) and (ii) resulting equations are much
easier for numerical implementation. In terms of the particle- and hole-like
amplitudes, the Anderson approximation reads
\begin{equation}
u_{i{\bf k}}({\bf r})={\cal U}_{ik}\psi_{i{\bf k}}({\bf r}), \quad
v_{i{\bf k}}({\bf r})={\cal V}_{ik}\psi_{i{\bf k}}({\bf r}),
\label{uv}
\end{equation}
where ${\cal U}_{ik}$ and ${\cal V}_{ik}$ are multiplicative factors (real) and
$\psi_{i{\bf k}}({\bf r})$ is the single-electron wave function, i.e.,
$\hat{H}_e\psi_{i{\bf k}}({\bf r})=\xi_{ik}\psi_{i{\bf k}}({\bf r})$. In the
case of interest, i.e., for a superconducting slab, we have
\begin{equation}
\psi_{i{\bf k}}({\bf r})=\frac{\varphi_i(z)}{\sqrt{L_xL_y}} e^{i(k_x x + k_y
y)},
\label{phi}
\end{equation}
where $\varphi_i(z)$ is the QWS wave function and $L_x,L_y$ are the dimensions
of the unit cell used for the periodic boundary conditions in the $x$ and $y$
directions. For infinite confinement of electrons in the slab with thickness
$d$~($L_x,L_y \gg d$) we have $\varphi_i(z)=\sqrt{2/d}\sin(\pi(i+1)z/d)$ and
\begin{equation}
\xi_{ik}=\frac{\hbar^2}{2m_{e}}\Big[\frac{\pi^2(i+1)^2}{d^{2}} + k^2\Big]-\mu.
\label{ksi}
\end{equation}
Inserting Eq.~(\ref{uv}) into Eq.~(\ref{bdg}) and introducing the
subband-dependent gap [in the case of interest $\Delta({\bf r})=\Delta(z)$]
\begin{equation}
\Delta_i=\int\!{\rm d}z\,\varphi^2_i(z)\Delta(z),
\end{equation}
we find a homogeneous system of two linear equations controlling ${\cal U}_{ik}$
and ${\cal V}_{ik}$. The determinant of the corresponding matrix should be equal
to zero for a nontrivial solution, leading to
$E_{ik}=\sqrt{\xi_{ik}^2+\Delta_i^2}$. Then, by using the standard normalization
condition ${\cal U}_{ik}^2 + {\cal V}_{ik}^2=1$, the BdG equations are reduced
to the BCS-like self-consistency equation
\begin{equation}\label{changedCP}
\Delta_{i^{\prime}}=\frac{1}{2}\int\!\!\frac{{\rm d}^2k}{(2\pi)^2}
\sum\limits_i\Phi_{i^{\prime}i}\frac{\Delta_i}{E_{ik}}\tanh(\beta E_{ik}/2),
\end{equation}
with the interaction-matrix element
\begin{equation}
\Phi_{i^{\prime}i} = g\int\!{\rm d}z\,\varphi^2_{i^{\prime}}(z) \varphi^2_i(z).
\label{intPhi}
\end{equation}
As $T$ approaches the critical temperature $T_c$, $E_{ik}$ reduces to the
single-electron energy $\xi_{ik}$ independent of $\Delta_i$. Hence, at $T_c$
Eq.~(\ref{changedCP}) becomes a linear homogeneous system of equations for
$\Delta_i$ that has a nontrivial solution when the corresponding determinant is
zero, i.e.,
\begin{equation}
\det[M_{i^{\prime}i}-\delta_{i^{\prime}i}]=0,
\label{Tc}
\end{equation}
from which we obtain $T_c$. In Eq.~(\ref{Tc})
\begin{equation}
M_{i^{\prime}i}
=\frac{m_e}{2\pi\hbar^2}\!\!\int\limits_{-\hbar\omega_D}^{\hbar\omega_D}\!\!\!{
\rm d}\xi\;\theta(\xi-\varepsilon_i)\frac{\Phi_{ii^{\prime}}}{2|\xi|}
\tanh\big(\beta_c|\xi|/2\big),
\label{TcA}
\end{equation}
with $\beta_c=1/(k_BT_c)$ and $\varepsilon_i=\frac{\hbar^2}{2m_{e}}
\frac{\pi^2(i+1)^2}{d^{2}}-\mu$. We note that a numerical solution of
Eq.~(\ref{Tc}) is significantly faster than the procedure based on a numerical
study of the BdG equations. The point is that Eq.~(\ref{Tc}) yields directly the
value of $T_c$. In contrast, to calculate $T_c$ from the BdG equations, one
first needs to find $\Delta_i$ as function of temperature. Then, $T_c$ can be
extracted from these data as the temperature above which there exists only zero
solution for $\Delta_i$. In addition, such a BdG-based numerical procedure of
determining $T_c$ is rather capricious in practice because at temperatures close
to $T_c$ convergence becomes rather slow and it is difficult to estimate an
appropriate number of iterations to approach a solution. This often results in
an overestimation of $T_c$ that is larger than the corrections to the Anderson
approximation, i.e., one more solid argument in favor of the present numerical
scheme.

\subsection{Quantum-size variations of the electron-electron coupling}
\label{formB}

In Sec.~\ref{formA} it is assumed that the phonon mediated coupling of electrons
$g$ is not position dependent. It is true for bulk and a good approximation even
for metallic nanofilms with thickness $d \gtrsim 10$-$20\,{\rm nm}$~(see our
results below). However, this is not the case for ultrathin nanofilms due to
significantly different (as compared to bulk) conditions for the lattice
vibrations at the interface between the film and the semiconductor substrate. As
we are interested in superconducting quantum-size oscillations typical for
ultrathin metallic films, the question arises how our formalism should be
modified to take account of this feature.

The coupling between electrons will depend on the proximity to the interface,
i.e., $g=g(z)$. Deep in the film, $g(z)$ approaches the bulk coupling $g_0$. For
${\rm Pb}$ we take~\cite{degen,fett} $g_{0}N_{\rm bulk}(0)=0.39$, with $N_{\rm
bulk}(0)$ the bulk density of states at the Fermi energy. Approaching the
interface, $g(z)$ is not equal to $g_0$ any longer but acquires a different
value $g_{\rm if}$. Therefore, to first approximation we can replace $g(z)$ by
the step function
\begin{equation}
g(z)=\left\{
\begin{array}{c}
g_{\rm if}\quad 0\leq z \leq d_{\rm if},\\
g_0\quad d_{\rm if} < z \leq d,
\end{array}
\right.
\label{gz}
\end{equation}
where $d_{\rm if}$ can be interpreted as the interface thickness. When
introducing the spatial dependence of the coupling between electrons,
Eq.~(\ref{intPhi}) is replaced by
\begin{equation}
\Phi_{i^{\prime}i} = \int\!{\rm d}z\, g(z)\, \varphi^2_{i^{\prime}}(z)
\varphi^2_i(z),
\label{intPhiz}
\end{equation}
so that we obtain
\begin{equation}
\Phi_{i^{\prime}i}\approx\frac{\bar{g}}{d}(1+\delta_{i^{\prime}i}),
\quad\bar{g}\approx\frac{1}{d}\int\limits_0^d{\rm d}zg(z).
\label{barg}
\end{equation}
As seen, to take into account the interface effect on the electron-electron
effective coupling, one should replace $g$ in all the relevant formulas in
Sec.~\ref{formA} by its spatially-averaged value $\bar{g}$ given by
Eq.~(\ref{barg}). Based on Eqs.~(\ref{gz}) and (\ref{barg}), we can find
\begin{equation}
\bar{g} = g_0 -\frac{d_{\rm if}(g_0-g_{\rm if})}{aN},
\label{barg}
\end{equation}
where $a$ is the lattice constant in the direction of the film growth and $N$ is
the number of monolayers in the film ($d=aN$, and we take $a=0.286 \AA$ for
${\rm Pb}(111)$, see Ref.~\onlinecite{PhysRevB.66.233408}). As seen from
Eq.~(\ref{barg}), the presence of the interface results in an extra contribution
to $\bar{g}$ decaying with increasing thickness as $1/N$. This is typical of the
surface effects whose relative contribution can be estimated as roughly
proportional to the ratio of the whole surface area to the system volume. Such a
contribution to the coupling was first reported in
Ref.~\onlinecite{PhysRevLett.88.256802}, where the electron-phonon mass
enhancement parameter $\lambda$ in ${\rm Ag}$ films grown on top of ${\rm
Fe}(100)$ was experimentally studied and found to dependent on the film
thickness. As known, $\lambda = gN(0)$ for $g\to 0$, with $N(0)$ the density of
states, and $\lambda$ is often called the (dimensionless) coupling constant.
{\it Ab initio} DFT-calculations of $\lambda$ also revealed the presence
of the $1/N$-contribution, see, e.g.,
Refs.~\onlinecite{PhysRevLett.88.256802,PhysRevB.81.214519}. Note that the
well-known softening of the phonon modes due to surface effects (see, e.g.,
Ref.~\onlinecite{PhysRevB.7.3028}) can be incorporated in a similar manner.
Below we assume that all surface effects are included in $g_{\rm if}$. The same
holds for effects due to a protective cover.

While $g_0$ is the film-material constant, $g_{\rm if}$ is not and changes from
one experiment to another. Moreover, $g_{\rm if}$ varies also with changing
nanofilm thickness because of quantum-size oscillations in the interface
properties, see, for instance, theoretical results and discussion about the
surface energy and the work function in nanofilms.~\cite{PhysRevB.66.233408}
Generally, the quantum-size oscillations are controlled by a multiple-subband
structure appearing in the single-electron spectrum due to the formation of QWS.
Each time when the bottom of a subband (QWS) crosses $\mu$, a shape
(quantum-size) resonance appears~\cite{PhysRevLett.10.332} with a significant
effect on the system characteristics. From Eq.~(\ref{ksi}) it is clear that the
shape resonances are periodic with period half the Fermi wavelength
$\lambda_F/2$. Thus, assuming that $g_{\rm if}$ exhibits thickness-dependent
oscillations with period $\lambda_F/2$, Eq.~(\ref{barg}) can be represented in
the form
\begin{equation}
\bar{g}=g_0 -\frac{g_1(4\pi aN/\lambda_F)}{N},
\label{barg_os}
\end{equation}
where $g_1(x)$ is a periodic function with $g_1(0)=g_1(2\pi)$. Below, in our
numerical study of the BdG equations, we replace $g$ in the formulas of
Sec.~\ref{formA} by $\bar{g}$ given by Eq.~(\ref{barg_os}).

It is worth noting that one should distinguish oscillations of $g$ from those of
the electron-phonon mass enhancement parameter $\lambda$. As $\lambda \approx
gN(0)$, there are two sources of quantum-size oscillations in $\lambda$, i.e.,
(i) the density of states that exhibits an increase each time when a new QWS
crosses $\mu$, and (ii) the phonon-mediated coupling $g$. The oscillations of
$\lambda$ are a known point that was investigated previously in experiments and
in theoretical studies, see, e.g., Refs.~\onlinecite{PhysRevLett.88.256802,
PhysRevB.66.233408, PhysRevLett.95.096802, PhysRevB.81.214519}. However, effects
of the interplay of oscillations of $N(0)$ with those of $g$ were not
investigated and reported.

Experimentally measured values of the period of the quantum-size oscillations in
${\rm Pb}$ nanofilms are found to be very close to $2\,{\rm ML}$. However, the
period is not exactly $2\,{\rm ML}$ so that the even-odd staggering in the basic
superconducting properties has phase-slip points where the phase of the even-odd
oscillations changes by $\pi$. When $4a/\lambda_F=1$, i.e., the period is equal
to $2\,{\rm ML}$, then we deal only with two parameters $g_1(\pi)$ and
$g_1(2\pi)$ because $g_1(4\pi aN/\lambda_F)$ is equal to $g_1(\pi)$ for an odd
number of monolayers and to $g_1(2\pi)$ for an even number of ML's. On the
contrary, when $4a/\lambda_F$ is not an integer, we need to go into more detail
about the dependence of $g_1$ on $N$. There is the possibility to overcome this
problem by using, say, a ``minimal scenario" when choosing the chemical
potential such that the period of the quantum-size oscillations is exactly
$2\,{\rm ML}$. In this case we are able to limit ourselves to only two
fitting parameters $g_1(\pi)$ and $g_1(2\pi)$. Therefore, our theoretical
analysis will be restricted to the data between the phase-slip points where the
effects of the mismatch between $\lambda_F/2$ and $2\,{\rm ML}$ can be neglected.
Within such a minimal scenario, the parabolic band approximation used
in Eq.~(\ref{ksi}) dictates that $\mu=1.150\,{\rm eV}$. It corresponds to
$\lambda_F$ being four times the single monolayer thickness $a$, i.e.,
$\lambda_F=1.14\,{\rm nm}$.

Now, we have everything at our disposal to check if the interplay between
thickness-dependent oscillations of $N(0)$ and $g$ can provide a solid
understanding of the significant variations in the amplitude of the quantum-size
oscillations of $T_c$ found in different experiments.

\section{Results and discussions}\label{result}
\begin{figure}[t]
\includegraphics[width=1.0\linewidth]{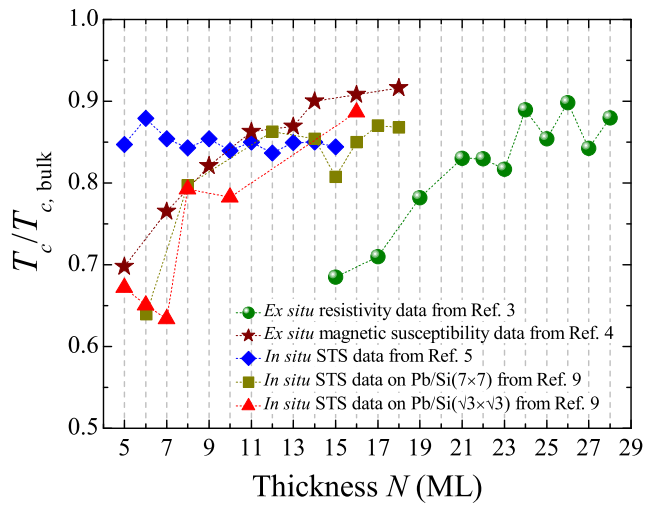}
\caption{(Color online) Experimental data on the thickness dependence of the
critical temperature $T_c$ of ${\rm Pb}$ nanofilms on ${\rm Si}(111)$ from
different experiments.} \label{fig1}
\end{figure}

\subsection{Brief review of experimental results on the $T_c$-oscillations}
\label{brief}

In Fig.~\ref{fig1} we show all reported sets of recent experimental data on the
thickness dependence of $T_c$ in ultrathin ${\rm Pb}(111)$ nanofilms on silicon.
The data indicated by circles are from Ref.~\onlinecite{science306_1915} and
were obtained by {\it ex situ} resistivity measurements. Here ${\rm Pb}$
nanofilms were deposited on a ${\rm Si}(111)$ substrate (with the ${\rm
Si}(111)\,(7\times7)-$reconstruction) and covered by a
protective ${\rm Au}$ layer with thickness $4\,{\rm ML}$. This set of data
exhibits clear even-odd quantum-size staggering for thicknesses of
$22$-$28\,{\rm ML}$. For smaller thicknesses there are only data for odd number
of atomic layers, i.e., $N=15,\,17,\,19$ and $21$. A clear phase slip of $\pi$
in the even-odd oscillations can be observed at $N=22\,{\rm ML}$. At this point
the even-odd staggering in $T_c$ changes its trend: above the thickness
$N=22\,{\rm ML}$ the critical temperature increases when passing from an odd
number of monolayers to an even number, however it decreases from $N=21\,{\rm
ML}$ to $N=22\,{\rm ML}$. Below we will limit the analysis of this data-set to
$N \geq 23$.

The results of Ref.~\onlinecite{nature2_173} are given by stars and obtained
from {\it ex situ} magnetic-susceptibility measurements for ${\rm Pb}$ films on
${\rm Si}(111)$ with the interface ${\rm Si}(111)(\sqrt{3}\times\sqrt{3})$. The
samples were protected by a ${\rm Ge}$ cap. The set includes data for films with
an odd number of atomic layers $N=5,\,7,\,9,\,11,\,13$ and with even numbers $N
= 14,\,16$ and $18$. Here it is not possible to obtain significant information
about the even-odd oscillations of $T_c$: we can only compare the data for
$N=13$ and $N=14$. As seen from this comparison, the increase of $T_c$ from
$N=13$ to $N=14$ is about two times smaller than the averaged amplitude of the
even-odd oscillations observed in the previous set of data from
Ref.~\onlinecite{science306_1915}.

The third experimental set of thickness dependent $T_c$ (rhombuses,
Ref.~\onlinecite{PhysRevLett.96.027005}) is for ${\rm Pb}$ nanofilms on the
substrate ${\rm Si}(111)(7\times7)$. Here the data were measured by
{\it in situ} scanning tunneling spectroscopy. There was no protective layer
covering the ${\rm Pb}$ film in this case. Clear signatures of even-odd
quantum-size oscillations are visible down to thickness of $5\,{\rm ML}$.
However, the amplitude of these oscillations is about $2$-$3$ times smaller than
that of the first data-set.~\cite{science306_1915} It is worth noting that $T_c$
of the third data set does not show an overall decrease with decreasing film
thickness. It is seen that there are two points where the phase of the even-odd
staggering of $T_c$ slips by $\pi$, i.e., at $N=6$ and at $N=14$. For this set
of data we will focus on the domain $7\leq N\leq 13$.

\begin{figure*}
\includegraphics[width=0.7\linewidth]{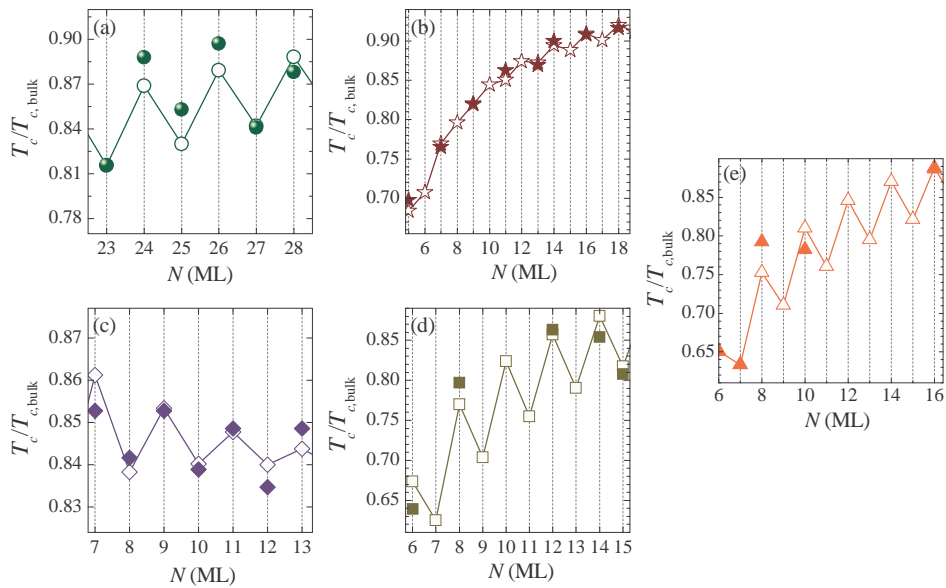}
\caption{(Color online) Theoretical results for the even-odd oscillations of
$T_c$~(open symbols) versus the experimental data (solid symbols): (a)
theoretical data were calculated for $g_1(\pi)=1.67g_0$~(odd numbers) and
$g_1(2\pi)=2.13g_0$, the experimental set is from
Ref.~\onlinecite{science306_1915}; (b) theoretical data were calculated
for $g_1(\pi)=0.64g_0$~(odd) and $g_1(2\pi)=1.46g_0$~(even), experimental
results are from Ref.~\onlinecite{nature2_173}; (c) the same as in the
previous panels but for $g_1(\pi)=-0.12g_{\rm eff}$ and $g_1(2\pi)=0.84g_{\rm
eff}$~(where $g_{eff}N_{\rm bulk}(0)=0.36$) and the experimental set is from
Ref.~\onlinecite{PhysRevLett.96.027005};
(d) the same as in previous panels but for $g_1(\pi)=1.08g_0$ and
$g_1(2\pi)=1.54g_0$ and the experimental set is from
Ref.~\onlinecite{PhysRevLett.102.207002} (for the interface ${\rm
Si}(111)(7\times7)-{\rm Pb}$); (e) the theoretical data were calculated for
$g_1(\pi)=1.05g_0$ and $g_1(2\pi)=1.59g_0$ and the experimental set is from
Ref.~\onlinecite{PhysRevLett.102.207002} (for the interface ${\rm
Si}(111)(\sqrt{3}\times\sqrt{3})-{\rm Pb}$).} \label{fig2}
\end{figure*}

The fourth (squares) and fifth (triangles) data sets are from the same
group~\cite{PhysRevLett.102.207002} and were measured by {\it in situ} scanning
tunneling spectroscopy. They correspond to ${\rm Pb}$ nanofilms on a
${\rm Si}(111)$ substrate with reconstruction ${\rm Si}(111)(7 \times 7)$ and
${\rm Si}(111)(\sqrt{3}\times\sqrt{3})$, respectively. The fourth set shows
the even-odd staggering of $T_c$ for $N=14$-$18$. However, here the slip-point
is situated at $N=16$, which significantly reduces the number of points
available for our analysis~(below we work with data for $N \leq 15$). The fifth
experimental set for $T_c$ does not provide us with a clear pattern of even-odd
oscillations. Only from the data for film thicknesses $N=5,\,6$ and $7$ we can
guess that the amplitude of such oscillations is not pronounced. The phase-slip
point of the even-odd staggering is visible at $N=6$~(below we make a numerical
fit to the data for $N \geq 7$). We note that the data of the fourth set are in
significant variance with those of the third set in spite of the fact that the
type of measurement and the substrate-film interface were similar in both
experiments.

\subsection{Impact of the thickness-dependent coupling between electrons}
\label{impact}

Results of our numerical analysis (open symbols) are given versus the
experimental data (solid symbols) in Fig.~\ref{fig2}. In panel (a) we compare
our data with the $T_c$-values from the first set
(Ref.~\onlinecite{science306_1915}). Here we choose the range $N=23$-$28$, i.e.,
above $N=22$, the experimental phase-slip point of the even-odd oscillations in
$T_c$~(see the discussion in the previous section). Our calculations have been
performed for $g_1(\pi)=1.67g_0$~(odd numbers) and $g_1(2\pi)=2.13g_0$~(even
numbers), which yields good agreement with experiment. At first sight these
values of $g_1$ seem too large to produce a deviation of only about
$10\%$-$20\%$ from bulk. However, $g_1/N$ is the correction to the
electron-electron coupling rather than $g_1$. In particular, for $N=20$ we
obtain $g_1/N=0.084g_0$ for odd numbers and $g_1/N=0.11g_0$ for even numbers. It
is worth noting that $g_1$ is larger for even numbers and so the resulting
coupling $g_0-g_1/N$ is smaller. However, $T_c$ is generally larger for even
numbers as compared to that for odd numbers in Fig.~\ref{fig1}(a). The reason
for this counterintuitive behavior is that there is an essential interplay
between the even-odd oscillations in the density of states $N(0)$ and similar
oscillations in the coupling $g$. In the case of panel (a) the drop in $g$ for
even $N$ is fully overcome by an increase in $N(0)$ so that the product
$gN(0)$~[$T_c \sim e^{-1/(gN(0))}$] slightly increases for even-number films and
decreases for odd-number films.

Figure \ref{fig2}(b) shows our theoretical results versus the experimental data
of the second set from Ref.~\onlinecite{nature2_173}. Here we used the fitting
parameters $g_1(\pi)=0.64g_0$~(for odd numbers) and $g_1(2\pi)=1.46g_0$~(for
even numbers). As seen, $g_1$ for odd numbers is three times smaller than for
even numbers but $T_c$ does not exhibit any pronounced even-odd staggering. The
explanation is again the interplay of the quantum-size variations of $N(0)$ with
thickness-dependent oscillations in $g$: the effect of the difference between
$g_1(\pi)$ and $g_1(2\pi)$ is almost compensated due to significant drops of
$N(0)$ for even numbers of atomic layers. It is worth noting the surprisingly
good agreement over the full range of experimental data.

Now we switch to the analysis of the third set of experimental data for $T_c$
from Ref.~\onlinecite{PhysRevLett.96.027005}. These data are compared with our
theoretical results in Fig.~\ref{fig2}(c) for $N=7\,{\rm ML}$ to $N=13\,{\rm
ML}$~(i.e., between the two phase-slip points $N=6$ and $N=14$). For this
data-set $T_c$ does not exhibit an overall decrease with decreasing $N$ but
oscillates around a value that is about $15\%$ lower than the bulk critical
temperature of $7.2\,{\rm K}$. Even for $N=50\,{\rm ML}$ the critical
temperature was found to be about $15\%$ smaller than in bulk. The bulk value of
$T_c$ and the superconducting energy gap were achieved only for relatively thick
films with $N=500\,{\rm ML}$.  To incorporate this feature, we introduce a
slightly smaller effective electron-phonon coupling constant $g_{\rm eff}$ which
replaces the bulk coupling constant $g_0$. By choosing $g_{\rm eff}N_{\rm
bulk}(0)=0.36$, $g_1(\pi)=-0.12g_{\rm eff}$ and $g_1(2\pi)=0.84g_{\rm eff}$, we
compare our results (the open symbols) with the experimental critical
temperature data (the solid symbols). As follows from Fig.~\ref{fig2}(c), we
again obtain very good agreement with the experimental data. As $g_1(\pi) <
0$, the total electron-phonon coupling constant $g_{\rm eff} - g_1/N$ for the
odd-layered nanofilms is slightly above the effective-bulk value $0.36$ and
increases when decreasing $N$. This is totally different from our analysis of
the previous data sets, and the direct consequence of this feature is that $T_c$
does not show an overall decrease with decreasing nanofilm thickness. As seen,
in this case the thickness-dependent oscillations of the electron-electron
coupling ``kills" such a decrease typical of the remaining experimental data
sets. Note that the appearance of the positive interface-induced contribution to
$g$ for the odd-layered films, is not a surprise. For example, a similar effect
was found for ${\rm Ag}$ films deposited on top of ${\rm Fe}(100)$ whisker.

Figures~\ref{fig2}(d) and (e) show the comparison of our numerical data with
data (the fourth and fifth sets, respectively) for $T_c$ from
Ref.~\onlinecite{PhysRevLett.102.207002}. For  ${\rm Pb}$ nanofilms grown
epitaxially on ${\rm Si}(111)(7\times 7)$~[panel (d)] we performed calculations
with the fitting parameters $g_1(\pi)=1.08g_0$ and $g_1(2\pi)=1.54g_0$. For
films with interface ${\rm Si}(111)(\sqrt{3}\times\sqrt{3})-{\rm Pb}$ we used
$g_1(\pi)=1.05g_0$ and $g_1(2\pi)=1.59g_0$. For the fourth set we focus on the
range below the phase slip point $N=16$~($N \leq 15$). For the fifth set we work
above the phase-slip thickness $N=6$~($N \geq 7$). Here $T_c$ increases for the
even-layered nanofilms and decreases for the odd-layered ones for both data
sets. However, similar to what was previously found for the first data set, the
coupling $g_0-g_1/N$ exhibits just the opposite size oscillations. As $T_c$
follows the trend $\sim e^{-1/(gN(0))}$, one sees that the thickness-dependent
oscillations of $N(0)$ are $\pi$-shifted with respect to the oscillations of
$g$, but the increase of $N(0)$ has a more significant effect
than the corresponding decreases of $g$ for even number of MLs. It is
interesting to note that the fourth and fifth experimental data-sets are
characterized by almost the same coupling between electrons (the difference in
$g$ is less than $1\%$). So, for the experiments reported in
Ref.~\onlinecite{PhysRevLett.102.207002}, $g$ is not sensitive to the change of
${\rm Si}(111)$ surface crystal ordering from $(7\times7)$ to
$(\sqrt{3}\times\sqrt{3})$. Here the question arises why the results of
Ref.~\onlinecite{PhysRevLett.102.207002} are so different with respect to those
of Ref.~\onlinecite{PhysRevLett.96.027005}. One possible reason may be the
difference in the quality of the silicon substrate which may have a significant
impact on the interface phonons, as seen form the completely different values of
$g_1$ found from our analysis in (c) and (d).

Note that the first and second data-sets in the present study were considered in
our previous work.~\cite{PhysRevB.75.014519} However, in that paper we performed
time consuming calculations invoking the full BdG formalism. As already
explained above (see the discussion after Eq.~(\ref{Tc})), the full-BdG-based
numerical procedure of determining the critical temperature often results in
an overestimation of $T_c$ that are larger than the errors of about
a few percent induced by the Anderson approximation. This is reason why in
the present paper we have chosen to use the Anderson-approximation-based
analysis of the available data, including the two data-sets considered
previously in Ref.~\onlinecite{PhysRevB.75.014519}.

We also note that we did not discuss error bars for the experimental data in the
present paper because they do not influence our results qualitatively: the
fitting parameters may be slightly different but the main conclusion about the
crucial importance of the interplay of quantum-size oscillations of $g$ and
$N(0)$ will remain unaltered.

\section{Conclusions}
\label{conclusions}

We have analyzed five experimental data-sets for the critical temperature of
ultrathin ${\rm Pb}$ nanofilms deposited on top of ${\rm Si}(111)$. We have
demonstrated that all these data sets can be reproduced within the BdG
formalism. We addressed the important problem that the
experimentally found amplitudes of the quantum-size oscillations of $T_c$ vary
significantly from one experiment to another. We showed that the reason for this
difference is an interplay of the quantum-size variations in the single-electron
density of states with the thickness-dependent oscillations in the
phonon-mediated coupling between the electrons. Such oscillations of the
coupling depends on the substrate material, the quality of the interface, the
presence or absence of a protection cover and other fabrication details that
change from one experiment to another. In addition, we demonstrated that
oscillations in the coupling $g$ and their interplay with the
thickness-dependent variation of the density of states
can even significantly change the overall trend of $T_c$ with decreasing
nanofilm thickness.

\begin{acknowledgments}
This work was supported by the Flemish Science Foundation (FWO-Vl).
\end{acknowledgments}

\end{document}